\documentclass[preprint,12pt]{elsarticle}
\usepackage{graphicx}
\usepackage{xcolor}
\usepackage{amsmath}
\usepackage{comment}
\usepackage{listings}
 \usepackage[utf8]{inputenc} 
\usepackage{hyperref} 
\usepackage{algorithm}
\usepackage{algorithmic}
\usepackage{ulem}
\makeatletter
\providecommand{\doi}[1]{%
  \begingroup
    \let\bibinfo\@secondoftwo
    \urlstyle{rm}%
    \href{http://dx.doi.org/#1}{%
      \discretionary{}{}{}%
      \nolinkurl{#1}%
    }%
  \endgroup
}
\makeatother

\usepackage{graphicx}
\usepackage{amssymb}

\usepackage{lineno}

\usepackage{url}

\usepackage{listings}
\usepackage{xcolor}

\definecolor{codegreen}{rgb}{0,0.6,0}
\definecolor{codegray}{rgb}{0.5,0.5,0.5}
\definecolor{codepurple}{rgb}{0.58,0,0.82}
\definecolor{backcolour}{rgb}{0.95,0.95,0.92}

\lstdefinestyle{mystyle}{
    backgroundcolor=\color{backcolour},   
    commentstyle=\color{codegreen},
    keywordstyle=\color{magenta},
    numberstyle=\tiny\color{codegray},
    stringstyle=\color{codepurple},
    basicstyle=\ttfamily\footnotesize,
    breakatwhitespace=false,         
    breaklines=true,                 
    captionpos=b,                    
    keepspaces=true,                 
    numbers=left,                    
    numbersep=5pt,                  
    showspaces=false,                
    showstringspaces=false,
    showtabs=false,                  
    tabsize=2
}

\journal{Computer Physics Communication}

\begin{document}

\begin{frontmatter}


\title{TensorBNN: Bayesian Inference for Neural Networks 
using
TensorFlow}



\author[Davidson]{B.S. Kronheim\fnref{alt1}}
\ead{brkronheim@davidson.edu}

\author[Davidson]{M.P. Kuchera}
\ead{mikuchera@davidson.edu}
\author[FSU]{H.B. Prosper}
\address[Davidson]{Department of Physics, Davidson College, Davidson, NC 28036}
\address[FSU]{Department of Physics, Florida State University, Tallahassee, FL 32306}
\fntext[alt1]{Current address: {\it Department of Physics, University of Maryland, College Park, MD, 20742}}
\begin{abstract}

\texttt{TensorBNN} is a new package based on  \texttt{TensorFlow} that implements Bayesian inference for modern neural network models. The posterior density of neural network model parameters is represented as a point cloud sampled using Hamiltonian Monte Carlo. The \texttt{TensorBNN} package leverages \texttt{TensorFlow}'s architecture and training features as well as its ability to use  modern graphics processing units in both the training and prediction stages.
\end{abstract}

\begin{keyword}
Bayesian Neural Networks \sep Machine Learning \sep \texttt{TensorFlow} \sep Hamiltonian Monte Carlo \sep arXiv: 2009.14393
\PACS 02.50.Tt

\end{keyword}
\end{frontmatter}



\label{S:1}
\section{Introduction}
\label{sec:TensorBNN}
\texttt{TensorBNN} is a flexible implementation of Bayesian neural networks (BNNs) built with \texttt{TensorFlow} \cite{tensorflow2015-whitepaper} and \texttt{TensorFlow-Probability} (\texttt{TFP})~\cite{tfp}, a popular machine learning platform with efficient co-processor computations. This implementation takes a Monte Carlo approach to make Bayesian predictions, in contrast to many current gradient descent-based approaches.
 \color{black}

The Flexible Bayesian Modeling (\texttt{FBM}) toolkit, developed by Radford Neal\,\cite{neal_1996, neal}, provides extensive capabilities for Bayesian inference for neural networks. However, machine learning technologies have evolved significantly since the first release of \texttt{FBM}. Robust, flexible machine learning platforms such as Google's \texttt{TensorFlow} \cite{tensorflow2015-whitepaper} and Facebook's \texttt{PyTorch} \cite{NEURIPS2019_9015} contain functionality in architecture design and optimization methods that are unmatched in frameworks with smaller user  and development support. \color{black} 
In addition, these packages provide a seamless interface with  highly-parallel co-processors such as Graphics Processing Units (GPUs) and Tensor Processing Units (TPUs), enabling large-scale computations, often \color{black} with orders of magnitude speedup over CPU-only software. The \texttt{TensorBNN} package
leverages these recent developments in order to provide an environment 
for Bayesian inference using the methods proposed by Neal\,\cite{neal_1996}. 

Implementations of approximate BNNs, which
 train efficiently on co-processors, \color{black}
such as the \texttt{DenseFlipout} layers in \texttt{TFP} based on  \cite{wen2018flipout}\color{black}, have recently appeared. These \texttt{DenseFlipout} layers approximate the prior and posterior densities with explicit functional forms, such as a Gaussian density, and optimize the parameters these functional forms using gradient descent. While such methods can be effective and are much less computationally expensive than a full Bayesian analysis, they are limited by the choice of the functional form of the posterior densities and,
therefore, are unlikely to be able to capture the full complexity of the posterior density over the parameter space of a neural network. 

The \texttt{TensorBNN} package approximates
the posterior density of the parameters of a neural network as a point cloud, that is, a neural network model is ``trained" not by finding a single neural network that best fits the training data, but rather by creating an {\it ensemble} of neural networks by sampling their parameters from the posterior density using a Markov chain Monte Carlo (MCMC) method. 

The paper is organized as follows. We begin in Section\,\ref{sec:BNN} with a description of the salient mathematical underpinning of BNNs and \texttt{TensorBNN}, in particular. This is followed by a description of the  \texttt{TensorBNN} package in Section\,\ref{sec:imp}. In Section\,\ref{sec:usecases} the performance of \texttt{TensorBNN} is illustrated  with a simple example. A summary is given in Section\,\ref{sec:conclusion}.

\section{Bayesian neural networks}
\label{sec:BNN}

\subsection{Mathematical Details}

Neural networks are 
 widely used  {\it supervised} machine learning models in which the training data comprises known inputs and outputs from which regression or classification models can be 
constructed.
The standard approach to training such a model, that is, fitting the model to data, is to minimize a suitable empirical risk function, which in practice is proportional to the average of a loss function. If the average loss is a linear function of the negative log likelihood, $-\ln p(D \, |\, \theta)$ of the data $D$, then minimizing the average loss is identical to estimation of the neural network parameters $\theta$ via maximum likelihood. In this case, 
fitting a model, that is, a function $f(x, \theta)$\color{black}, to data can be construed as an {\it inference} problem.  Furthermore, if a prior density $\pi(\theta)$ over the parameter space can be defined, then the inference can be performed using the Bayesian approach\,\cite{neal_1996}. 
\color{black}

A Bayesian neural network can be represented as follows 
\begin{align}
    p(x, D) & = \int {f}(x, \theta)\color{black} \, p(\theta \, | \, D) \, d\theta ,
\label{eq:BNN}
\end{align}
where 
\begin{equation}
 p(\theta \, | \, D)  = \frac{p(D \, |\, \theta)\, \pi(\theta)}{p(D)}, 
 \label{eq:posterior}
 \end{equation}
 is the posterior density
over the parameter space $\Theta$ of the neural network
with parameters $\theta \in \Theta$. Since each set of parameters $\theta$ corresponds to a different neural network,  
each set of parameters \color{black} is associated with a different output value $y = f(x, \theta)$ for the same input $x$.  Therefore, the posterior density and neural network $f(x, \theta)$ together induce a
predictive distribution given by
\begin{align}
    p(y \, | \, x, D) & = \int \delta(y - f(x, \theta)) \, p(\theta \, | \, D) \, d\theta .
    \label{eq:predictive}
\end{align}
The quantity $D = \{(t_k, x_k)\}$ denotes the training data,
which consists of the targets $t_k$ associated with data $x_k$. 

In practice, the posterior
density is  approximated \color{black} by a point cloud $\{ \theta_i \}$
sampled using a Monte Carlo method
\color{black}
and Eq.\,(\ref{eq:predictive}) is approximated by binning the values $y = f(x, \theta)$ with $\theta \in \{\theta_i \}$.
A practical advantage of the
Monte Carlo
sampling approach is that it provides a straightforward method to estimate the uncertainty due to the finite size of the training data in any quantity that depends on the network parameters: one simply computes the quantity for every sampled parameter point which yields an ensemble of estimates.
\color{black}

\color{black}

The Monte Carlo sampling approach makes no assumptions about the form of the  posterior density  of the neural network parameters, whereas uncertainty quantification using methods such as the \texttt{DenseFlipout} \cite{wen2018flipout}  approximate the posterior density using a specific functional form.
However, the Bayesian approach requires the specification of a prior, which introduces its own
complication. \color{black} But, since the network parameters are sampled from the posterior density, the sensitivity of inferences to the choice
of prior can be assessed by re-weighting the sampled points by the ratio of the
new prior to the one with which the sample was generated. Moreover, this flexibility can even be used to optimize the choice of prior.

\color{black}
The current version of \texttt{TensorBNN} is restricted to fully-connected deep neural networks (DNN),
\begin{equation}
  f(\mathbf{x}, \theta) = g(\mathbf{b}_K + \mathbf{w}_K \, h_K(\cdots + h_1(\,\mathbf{b}_1 + \mathbf{w}_1 \, h_0(\mathbf{b}_0 + \mathbf{w}_0 \, \mathbf{x}) \, )\cdots). 
  \label{eq:dnn}
\end{equation}
The quantities $\mathbf{b}_j$ and $\mathbf{w}_j$---the biases and weights---are matrices of parameters and $h_j$ and $g$ are the activation and output functions, respectively; $j$ is the layer number.
\color{black} In
Eq.\,(\ref{eq:dnn}), the functions $h_j$ and $g$ are applied element-wise to their matrix arguments.
\texttt{TensorBNN} maintains  many of the standard activation function options available in \texttt{TensorFlow}, shown in Table\,\ref{tab:activationFunctions}, \color{black} with the addition of one custom activation function, a modified \texttt{PReLU} \cite{He2015DelvingDI} function, called \texttt{SquarePReLU}
\begin{equation}
    h_j(x) = 
    \begin{cases}
        x & \textrm{for }x\ge 0\\    
        a_j^2 x & \textrm{otherwise}.   
\end{cases}
\end{equation}
We opted to  choose the trainable \color{black} parameter to be $a_j$ rather than $a_j^2$ to ensure that the slope remains positive thereby keeping the activation function one-to-one. %

\begin{table}[ht]
\begin{center}
\begin{tabular}{ | p{2.5cm} | p{6cm} | } 
\hline
{\bf Name} & {\bf Description} \\ 
\hline 
\texttt{Sigmoid} 
&
$1 / (1+e^{-x})$ \vspace{0.05cm}\\
\hline
\texttt{Tanh} & $\tanh(x)$ 
\vspace{0.05cm}\\
\hline
\texttt{Softmax} & ${e^{x_i} / \sum_n e^{x_n}}$
\vspace{0.05cm}\\
\hline
\texttt{Exp} & $e^{x}$
\vspace{0.05cm}\\
\hline

\texttt{ReLU} & $ \begin{cases} 
      0 & x\leq 0 \\
      x &  x > 0 
   \end{cases}
$
\vspace{0.05cm}\\
\hline
\texttt{leakyReLU} & $ \begin{cases} 
      \alpha x & x\leq 0 \\
      x &  x > 0 
   \end{cases}
$
\vspace{0.05cm}\\
\hline
\texttt{Elu} & $ \begin{cases} 
      e^{x}-1 & x\leq 0 \\
      x &  x > 0 
   \end{cases}
$ 
\vspace{0.05cm}\\
\hline
\texttt{PReLU} & leakyReLU with trainable $\alpha$ \\
\hline
\texttt{SquarePReLU} &$ \begin{cases} 
      \alpha^2 x & x\leq 0 \\
      x &  x > 0 
   \end{cases}
$ $\alpha$ is trainable 
\vspace{0.05cm}\\
\hline
\end{tabular}
\end{center}
\caption{The built-in activation functions of \texttt{TensorBNN}.}
\label{tab:activationFunctions}
\end{table}

\color{black} The function $f$ is 
\begin{equation}
    f(x) = 
    \begin{cases}
    1/(1 + \exp(-x)) & \textrm{for a classifier and} \\
        x & \textrm{for regression models.}
\end{cases}
\end{equation}
 The weights and biases of the network along with  any parameters associated with the activation function are the free parameters of the neural network model. \color{black}

In the following subsection, we describe the pertinent details of this implementation, while the second subsection contains relevant information on the hardware used to perform the sampling for the examples described in Section\,\ref{sec:usecases}. 

\subsection{Likelihood and Prior}
\label{subsec:likelihood}
\paragraph{Likelihood} The likelihood functions included in the package are modeled as follows:
\begin{align}
    p(D \, | \, \theta) & = 
    \begin{cases}
        \prod_k y_k^{t_k} \, (1 - y_k)^{1 - t_k} & \textrm{for classification models and}\\    
        \prod_k \mathcal{N}(t_k, y_k, \sigma)  & \textrm{for regression models},    
\end{cases}
    \label{eq:likelihood}
\end{align}
where $y_k \equiv f(x_k, \theta)$ and $\mathcal{N}(x, \mu, \sigma)$ is a Gaussian density with mean $\mu$ and standard deviation $\sigma.$ \color{black}
For classifiers, the targets $t_k$ are 1 and 0 for true and false identification, while for regression models they are the desired regression values. 

\paragraph{Prior}
\color{black}

Constructing a prior for a multi-parameter likelihood function is exceedingly challenging and is especially so for functions as complicated as those in Eq.\,(\ref{eq:likelihood}). For an excellent and thorough review of this problem we recommend  Nalisnick\,\cite{eric_2018}). We sidestep this challenge by proceeding pragmatically and relying instead on {\it ex post facto} justification of the choices we make: the choices are justified based on the quality of the results. Moreover, \texttt{TensorBNN} includes a re-weighting mechanism for studying the sensitivity of the results to these choices.

Each layer of the DNN contains three kinds of parameter: weights, biases, and the slope parameters of the activation functions. In the \texttt{FBM} toolkit of  Neal\,\cite{neal_1996}, the prior for each weight and bias is a zero mean Gaussian density, with the precisions, $\sigma^{-2}$, of these densities constrained by gamma hyper-priors. 
In \texttt{TensorBNN}, there are several options. Specifically, users can choose between using Gaussian or Cauchy priors, and whether to use Gaussian hyper-priors or no hyper-priors. Using the default \texttt{GaussianDenseLayer}, the weights $w_i$ of a given layer are assigned the hierarchical
prior
\begin{align}
    \pi(\mathbf{w}) & =
    \left( \prod_i 
    \mathcal{N}(\omega_i, \alpha, \beta)\right) \\
    & \times 
    \mathcal{N}(\alpha, \mu_\alpha, \sigma_\alpha) \, \mathcal{N}(\beta, \mu_\beta, \sigma_\beta)
    & \textrm{(hyper-prior)},
    \label{eq:overallprior}
\end{align}
comprising a product of 
Gaussian densities with mean and standard deviation parameters, $\alpha$ and $\beta$, respectively, constrained by hyper-priors modeled as Gaussian densities. A hierarchical prior of the same form 
is assigned to the biases of a given layer, but with different mean and standard deviation parameters. The Gaussian density is chosen to constrain the  weights and biases within a reasonable range, with the option of effectively increasing the tails through the use of the hyper-priors. Using the Cauchy distributions would create even larger tails that accommodate a larger dynamic range in  the magnitude of the parameters,  while still enforcing the clustering of weights near zero. The hyper-priors were chosen to be Gaussian as the hyper-parameters are not expected to deviate greatly from their nominal values. 

\color{black}

The prior for the slope parameter of the activation function can be either an exponential density with its rate parameter constrained by an exponential hyper-prior 
\begin{align}
    \pi(m) & = 
    \lambda \exp(-\lambda m) \, 
    \gamma_\lambda \exp(-\gamma_\lambda \lambda),
    \label{eq:slopeprior1}
\end{align}
or a Gaussian prior with mean and
standard deviation parameters constrained
by Gaussian density hyper-priors as follows
\begin{align}
    \pi(m) & = 
    \mathcal{N}(m, \xi, \kappa) \,\mathcal{N}(\xi, \mu_\xi, \sigma_\xi) \, N(\omega, \mu_\kappa, \sigma_\kappa).
    \label{eq:slopeprior2}
\end{align}
 The slope parameter, $m$, should remain close to zero. To this end, we choose a exponential density, Eq.\,(\ref{eq:slopeprior1}), when this slope is positive, and a Gaussian density, Eq.\,(\ref{eq:slopeprior2}), when the slope can also be negative. 

\color{black}

The overall prior $\pi(\theta)$ is a product of the priors for all weights, biases, and slope parameters, and the associated hyper-priors. 
For regression models, there is an additional hyper-parameter, the standard deviation of the likelihood function given in Eq.\,(\ref{eq:likelihood}).
 The initial value of this hyper-parameter, which 
 can be \color{black} optimized during training, should reflect the user's best guess as to the approximate average error in the prediction, though an initial value of 0.1 is reasonable. 
See Table\,\ref{tab:priors} for a summary of the parameters in the priors.\color{black}

\begin{table}[ht]
\begin{center}
\begin{tabular}{ | p{2.5cm} | p{5cm} | }
\hline
{\bf Parameter} & {\bf Value}\\

\hline

$\beta$ & Gaussian standard deviation parameter\\
\hline
$\alpha$ & Gaussian mean parameter\\
\hline
\color{black}
$\mu_{\alpha}$ & mean of hyper-prior for $\alpha$ \\
\hline
$\sigma_{\alpha}$ & standard deviation of hyper-prior for $\alpha$\\
\hline
$\mu_{\beta}$ & mean of hyper-prior for $\beta$ \\
\hline
$\sigma_{\beta}$ & standard deviation of hyper-prior for $\beta$ \\
\hline
$\lambda$ & slope exponential distribution scale parameter \\
\hline
$\gamma_{\lambda}$ & scale parameter of hyper-prior for $\lambda$\\
\hline
$\xi$ &  slope Gaussian distribution mean parameter \\
\hline
$\kappa$ & slope Gaussian distribution standard deviation parameter \\
\hline
$\mu_{\xi}$ & mean of hyper-prior for $\xi$\\
\hline
$\sigma_{\xi}$ & standard deviation of hyper-prior for $\xi$ \\
\hline
$\mu_{\kappa}$ &  mean of hyper-prior for $\kappa$ \\
\hline
$\sigma_{\kappa}$ & standard deviation of hyper-prior for $\kappa$ \\
\hline
$\sigma$ & standard deviation of Gaussian likelihood function \\
\hline
\end{tabular}
\end{center}
\caption{\texttt{DenseLayer} initial values of prior hyper-parameters, $\alpha$ and $\beta$, and the fixed hyper-prior parameters.}
\label{tab:priors}
\color{black}
\end{table}
\color{black}

\subsection{Sampling the posterior density}
\label{sec:HMC}
 Since the high-dimensional integral in Eq.\,(\ref{eq:BNN}) is intractable it is typically approximated using  a Monte Carlo method to sample from the posterior density $p(\theta \, | \, D)$. The Monte Carlo method of choice is Hamiltonian Monte Carlo (HMC)\,\cite{neal_1996, Betancourt2017ACI} in which the posterior density is written as
$$p(\theta \, | \, D)= \exp(-V(\theta)),$$
where $V = -\ln{p(\theta \, | \, D)}$ is viewed as a ``potential'' to which a ``kinetic'' term $T = \mathbf{p}^2/2$ is added to form a ``Hamiltonian'' $H = T + V$. The dimensionality of the ``momentum'' $\mathbf{p}$ equals that of the parameter space $\Theta$. 
The HMC sampling algorithm alternates between deterministic traversals of the space $\Theta$ governed by a finite difference approximation of Hamilton's equations and random changes of direction. In order to achieve detailed balance and, therefore, ensure asymptotic convergence  to the correct posterior density,  the deterministic trajectories are computed using a reversible, leapfrog approximation, to
Hamilton's equations. 
The HMC method has two free parameters, the step size along the trajectories and the number of steps to take before executing a random change in direction. The method used to determine these parameters is described in Section\,\ref{subsec:adjust}.

\section{Implementation Details}
\label{sec:imp}

\texttt{TensorBNN}, which is built using the \texttt{TensorFlow}\,\cite{tensorflow2015-whitepaper} and \texttt{TensorFlow}-\texttt{Probability} \,\cite{tfp},  follows the design of BNNs described in \citet{neal_1996} with some improvements and modifications. Here we summarize the general structure of \texttt{TensorBNN} and the improvements, which  include the HMC parameter adaptation scheme and the addition of pre-training.

\subsection{Model Declaration}

\texttt{TensorBNN} provides a framework for the user-friendly construction of BNNs in a manner similar to the \texttt{Keras}\,\cite{chollet2015keras} interface for building neural networks with \texttt{TensorFlow}. The main object in the package is the \texttt{network} object,
which is the base for all operations in the package. The options that can be specified when instantiating this object are the data type, e.g. float32 or float64, the training and validation data, and the normalization scaling for the output. An example network declaration is shown below.
\color{black}
\begin{lstlisting}[language=Python, numbers=none]
model = network(dtype,     # Data type to use
                inputDims, # Length of 1 input
                trainX,    # Training input
                trainY,    # Training output
                validateX, # Validation input
                validateY) # Validation output
\end{lstlisting}

After the \texttt{network} object is instantiated, the layers and activation functions are added. Each of these is a variant of the \texttt{Layer} object. For example, a \texttt{GaussianDenseLayer} and a \texttt{CauchyDenseLayer} are included in the package with multiple activation functions, but users can create their own \texttt{Layer} variants and activation functions with custom priors. 

The layer objects can be initialized either randomly or with pre-trained weights and biases. When using random initialization, the Gaussian He \cite{He_2015_ICCV} initialization is used to determine starting values of the weights, and the biases are extracted from the same random distribution. This is done to keep the starting values small, while allowing variation. As discussed in Section\,\ref{subsec:likelihood}, the priors for the weights and biases are either Gaussian or Cauchy densities, with their parameters constrained by Gaussian hyper-priors (see Eq.\,(\ref{eq:overallprior})). 
These values, which are summarized in Table\,\ref{tab:denselayerconstants},cannot be changed within the layer objects, though it is a simple matter to create a new \texttt{Layer} variant.
\begin{table}[ht]
\begin{center}
\begin{tabular}{ | p{2.5cm} | p{2.5cm} | p{2.5cm} | }
\hline
{\bf Parameter} & {\bf Gaussian Value}& {\bf Cauchy Value}\\
\hline
$\alpha$ & 0 & 0 \\
\hline
$\beta$ & 1 & 0.5\\
\hline
\hline
$\mu_\alpha$ & 0 & 0\\
\hline
$\sigma_\alpha$ & 0.1 & 0.2\\
\hline
$\mu_\beta$ & 1 & 0.5 \\
\hline
$\sigma_\beta$ & 0.1 & 0.5\\
\hline
\end{tabular}
\end{center}
\caption{\texttt{GaussianDenseLayer} and \texttt{CauchyDenseLayer} initial values of prior hyper-parameters, $\alpha$ and $\beta$, and the fixed hyper-prior parameters.}
\label{tab:denselayerconstants}
\color{black}
\end{table}

In the code snippet below, a \texttt{DenseLayer} and an activation function are added to the network. This process would be the same for any layer or activation function with a different object added.
\begin{lstlisting}[language=Python,numbers=none]
model.add(            # Add layer command
        DenseLayer(   # Dense layer object
        inputDims,    # Size of layer input vector
        width,        # Size of layer output vector
        seed=seed,    # Random number seed
        dtype=dtype)) # Layer datatype
model.add(Tanh())     # Hyperbolic tangent activation
\end{lstlisting}

\color{black}

\begin{table}[ht]
\begin{center}
\begin{tabular}{ | p{2.6cm} | p{2.1cm} | p{7.4cm} | } 
\hline
{\bf Name} & {\bf Parameter} & {\bf Description} \\ 
\hline
\texttt{leakyReLU} & $m=0.3$ & fixed slope parameter \\
\hline
\texttt{PReLU} & $m=0.2$ & initial slope parameter \\
\hline
 & $\lambda$ = 0.3 & initial rate parameter for the $m$ prior\\
 \hline
 & $\gamma_\lambda = 0.3$ & rate parameter for the $\lambda$ hyper-prior\\
 \hline

\texttt{SquarePReLU} & $m=0.2$ & initial slope parameter \\
\hline
 & $\gamma$ = 0.3 & initial mean for the $m$ prior\\
 \hline
 & $\omega$ = 0.3 & initial standard deviation for the $m$ prior\\
 \hline
 & $\mu_\gamma$ = 0.0 & mean for the hyper-prior of $\gamma$\\
 \hline
 & $\sigma_\gamma$ = 0.1 & standard deviation for the $\gamma$ hyper-prior\\
 \hline
 & $\mu_\omega = 0.3$ & mean for the $\omega$ hyper-prior\\
 \hline
 & $\sigma_\omega$ = 0.3 & standard deviation for the $\omega$ hyper-prior\\
 \hline

\end{tabular}
\end{center}
\caption{The initial and fixed parameters of the built-in activation functions of \texttt{TensorBNN}.}
\label{tab:activationconstants} 
\end{table}

\color{black}
Within the package there are eight options for activation functions, which are listed in Table\,\ref{tab:activationFunctions}. All the activation functions except \texttt{SquarePReLU} are standard. The \texttt{SquarePReLU} was developed specifically for \texttt{TensorBNN}. Both \texttt{PReLU} and \texttt{SquarePReLU} have trainable slope parameters $\alpha$. They have, however, different priors and hyper-priors, given in Eqs.\,(\ref{eq:slopeprior1}) and (\ref{eq:slopeprior2}), with the fixed and initial values of their parameters listed in Table\,\ref{tab:activationconstants}. 
For \texttt{PReLU}, the exponential distribution was chosen to model the prior belief that the slopes should be close to zero. The rates were picked to allow for larger slopes, while still enforcing the belief that smaller slopes are preferred. 
\color{black}
For \texttt{SquarePReLU}, as we are considering the square root of the slope, which can be positive or negative, the Gaussian prior was chosen because it is continuous at 0 and enforces the prior preference for small slopes. 
Once again, the priors for these activation functions cannot be changed, but a custom activation function with different priors can be created.
When the \texttt{add} method of the \texttt{network} class is called, the layer or activation function is added to a list of layers. Additionally, the weights, biases, and trainable activation functions from the layer along with the hyper-parameters are stored.

\subsection{Hamiltonian Monte Carlo Initialization}

After building the network architecture, the Hamiltonian Monte Carlo (HMC) sampler must be initialized. This is done through a method of the \texttt{network} class. An example usage is presented below. As described in Section\,\ref{sec:HMC}, HMC is a Markov chain Monte Carlo method where sampling is performed by moving through the parameter space in a manner governed by a fictitious potential energy function determined by the posterior density of the neural network parameters. The numerical approximation used is the leapfrog method, in which the number of leapfrog steps together with the step size determine the distance traveled to the next proposed point.  Here leapfrog steps refers to the number of steps of the integrator for a single trajectory through the parameter space, leading to the next potential sample point. \color{black} Naturally, larger step sizes yield longer deterministic trajectories, but they also increase the accumulated error due to the numerical approximation and so lower the acceptance rate. Unfortunately, choosing good values for the number of steps and the step size can be challenging. Therefore, \texttt{TensorBNN} contains an algorithm,
called the parameter adapter, to find these parameters automatically.

The adapter searches a discrete space of step sizes and number of leapfrog steps using the algorithm of \cite{Wang:2013:AHR:3042817.3043100}. This discrete space is determined by a minimum and maximum step size and a minimum and maximum number of steps. In addition, the adapter accepts the number of iterations before a reset is performed. A reset is performed if after this number of iterations no point has been accepted. It is also given the number random pairs of step size and leapfrog steps to try before using the search algorithm from \cite{Wang:2013:AHR:3042817.3043100}.
Finally, it also accepts two constants $a$ and $\delta$, which assume the values 4 and 0.1, respectively as suggested in \cite{Wang:2013:AHR:3042817.3043100}. 
An example of the code used to initialize this is shown below. \color{black}

While the parameter adapter will be able to find a good choice in the region specified, if one exists,  some tuning is still needed for ideal parameter values. 
As discussed in \cite{pmlr-v139-izmailov21a}, a reasonable criterion for selecting an initial step size and the number of leapfrog steps is to require the total trajectory length, determined by the step size times the number of leapfrog steps, be roughly $\pi/2.$  \color{black}
Similarly, 
 the number of burn-in steps is a hyper-parameter of the algorithm.

\color{black} 
One of the changes from Neal's procedure is the use of HMC to sample the hyper-parameter space instead of Gibbs sampling. This was done because Gibbs sampling requires knowledge of the conditional distribution for each hyper-parameter given that the rest are fixed. While this is possible to calculate for some hyper-priors the package allows these to be changed to custom priors, for which it may not be possible to compute the conditional densities. It was, therefore, simpler to use a second HMC sampler to sample the hyper-parameters since this works for any hyper-prior. 
The number of steps for the HMC hyper-parameter sampler is kept constant, but the step size is modified  using the Dual-Averaging algorithm \cite{dualAveraging} \color{black} depending on the acceptance rate of the sample for 80$\%$ of the burn-in period. The \texttt{TensorFlow} - \texttt{Probability} HMC implementation is used for the sampling.

\begin{lstlisting}[language=Python,numbers=none]
model.setupMCMC(
        stepSizeStart=0.005,
        stepSizeMin=0.0025,
        stepSizeMax=0.01,
        stepSizeOptions=40, 
        leapfrogStart=2,
        leapfrogMin=2,
        leapfrogMax=50,
        leapfrogIncrement=1,
        hyperStep=0.01,
        hyperLeapfrog=5, 
        burnin=20,
        averagingSteps=2)
\end{lstlisting}

\subsection{Model Parameter Sampling}

The model parameters are sampled \color{black} using the the \texttt{train} method of the \texttt{network} class. The parameters of the \texttt{train} method determine 1) the likelihood function, 2) the metrics to be calculated during training, 
3) whether to use the hyper-priors, \color{black}  4) the number of burn-in epochs, 5) how often to save networks, and 6) the directory 
in which to store the models. An example is provided below. The prior is determined by the \texttt{Likelihood} object. The priors included in the package are those described in Section\,\ref{subsec:likelihood}, though any desired likelihood function can be used through a custom \texttt{Likelihood} object. 
The built-in Gaussian likelihoods used for regression allow specification of the initial value of the standard deviation. If \texttt{FixedGaussianLikelihood} is used the standard deviation will remain constant, but if \texttt{GaussianLikelihood} is used, the standard deviation will be treated like a hyper-parameter.  \color{black}

The metrics to be computed are specified by including a list of all desired \texttt{Metric} objects. The built-in metrics are \texttt{PercentError, SquaredError}, and \texttt{Accuracy}. All three accept the normalization scalars of  mean and standard deviation to calculate the metrics for unnormalized data, as well as an option to take the exponential of output values before computing for the metrics, for the case of log-scaled outputs. The \texttt{PercentError} metric is defined as
$$PE = 100\,\mathbb{E}\left[\frac{|y_{pred}-y_{true}|}{y_{real}}\right].$$
\texttt{Squared Error} is defined as 
$$SE = \mathbb{E}\left[(y_{pred}-y_{true})^2\right].$$
In both cases, the expectation is with respect to the input data, $y_{pred}$ is the predicted value for a given input, and $y_{true}$ is the target.
Finally, \texttt{Accuracy} is simply the number of correct predictions divided by the total number of predictions for a classification task.

\begin{lstlisting}[language=Python,numbers=none]
 # Declare Gaussian Likelihood with sd of 0.1
likelihood =  FixedGaussianLikelihood(sd=0.1)
metricList = [ # Declare metrics
    SquaredError(mean=0, sd=1, scaleExp=False),
    PercentError(mean=10, sd=2, scaleExp=True)]
network.train(
        320,   # Nmber of training epochs
        2,     # Increment between network saves
        likelihood,
        adjustHypers=True,
               # Whether to use the hyper-priors
               # or not
        metricList=metricList,
        folderName="Regression", 
               # Name of folder for saved networks
        networksPerFile=50)
               # Number of networks saved per file
\end{lstlisting}

In training, the HMC samplers run for the specified number of epochs. An epoch corresponds to iterating the differential equations inside the main HMC sampler for the specified number of leapfrog steps, updating the weights, biases, and activation functions, and then repeating this process with the hyper-parameters.

\subsection{Predictions and Post Processing}
 In order to prevent the output files in which the networks are saved from becoming too large and to allow predictions midway through training, 
 the number of networks written to each file 
 can be specified by the user, as noted above.
 One file is saved with the shapes of each matrix that defines the network architecture so that they can be properly extracted. Another file is saved that contains the layer names so the network can be reconstructed.

 The \texttt{predictor} object, which
 is instantiated as follows,
 \begin{lstlisting}[language=Python,numbers=none]
 # instantiate a predictor object
 network = predictor(
            "modelDir/", # Directory where model is located
            dtype=dtype) # Data type of model
 \end{lstlisting}
 uses these files to make predictions. Once the object is instantiated, predictions 
can be made by calling its \texttt{predict} method with an input data matrix and specifying  that every $n$ saved networks are to be used. 
\begin{lstlisting}[language=Python,numbers=none]
 
 predictions = network.predict(
                inputData,# Input data for model
                n=10)     # Predict using every 10 networks
 \end{lstlisting}

\color{black}
Beyond making predictions from saved networks, the \texttt{predictor} class is also capable of re-weighting networks given a new set of priors.  The ability to re-weight the point cloud of networks makes it possible to study the sensitivity of results to the choice of prior. We can compute the posterior density
$p_1(\theta|D)$ with the prior used in the generation of the point cloud and we can calculate the posterior density $p_2(\theta|D)$ using a different prior. By weighting each network $j$ by the ratio
$p_2(\theta_j|D)/p_1(\theta_j|D)$, given the network parameters $\theta_j$, we can approximate the point cloud that would have been generated had we used the alternative prior. In general, however, the only components of $p_1(\theta|D)$ and $p_2(\theta|D)$ that differ are their priors. Therefore, we can substitute $\pi_2(\theta_j)/\pi_1(\theta_j)$ for $p_2(\theta_j|D)/p_1(\theta_j|D).$ But since it may be useful to study the effect of different likelihoods on the results, the code includes the option to supply new likelihoods. The option to re-weight allows fine tuning of the networks after training and makes it possible to explore the impact of different priors on the results. 

Re-weighting is implemented in the \texttt{reweight} method
 of the \texttt{predictor}, which
requires an architecture file containing the architecture of the network with different priors. The method can also accept the training data and a \texttt{likelihood} object for use in calculating the network probabilities should the impact of the \texttt{likelihood} need to be studied. The user can choose to use only every $n$ networks when making  predictions
as illustrated in the code snippet below. 
In order to use these features a separate \texttt{Layer} object must be created for each new prior. These objects are then passed as a dictionary to the \texttt{predictor} object when it is instantiated. Optionally, a modified version of the \texttt{likelihood} object used while training can be included to study the impact of the modifications.
\color{black}
The code below shows the instantiation of a \texttt{predictor} object with a custom \texttt{Layer} added and a call of the \texttt{reweight} method which returns a weight for each network.

\begin{lstlisting}[language=Python,numbers=none]
 # declare predictor object
network = predictor("/path/to/saved/network",
                    dtype=dtype, 
                    # data type used by network
                    customLayerDict={"dense2": Dense2}
                    # A dense layer with a different 
                    # prior 
                    likelihood=modifiedLikelihood)
                    # An optional likelihood function
                    
weights = network.reweight(n=10,# Use every 10 saved networks
                    architecture="architecture2.txt")
                    # New architecture file
\end{lstlisting}

The \texttt{predictor} can also calculate the autocorrelation function of the networks, which is useful for choosing a suitable value for $n$ in the \texttt{predictor}. Given the sequence of networks $f_1, f_2, \cdots$, the  autocorrelation and
normalized autocorrelation are defined by
\begin{align}C(n) & = \frac{1}{N-n} \sum_{i=1}^{N-n} (f_{i} - \overline{f})(f_{i+n} - \overline{f}), \quad\textrm{and}\nonumber\\
   \rho(n) & = C(n) / C(0),
\label{eq:autocorrelation}
\end{align}
respectively, 
where $\bar{f}$ is the average prediction over the networks and $n$ is the autocorrelation lag. 
One expects an approximately exponential decrease of $\rho(n)$ with $n$. The larger the value of $n$ the smaller the correlation between the networks separated by a lag of $n$ and, therefore, the more independent the resulting ensemble of networks. 
The autocorrelation is computed with the \texttt{autocorrelation} method to which an input data matrix \texttt{inputData} is passed along with the maximum lag $n_\textrm{max}.$ 
The output of the method is a list containing the average autocorrelation for each value of $n$ from 1 to $n_\textrm{max}$, where $n$ enters the calculation as in Eq.\,(\ref{eq:autocorrelation}). The average is taken with respect to the input data
matrix. 

The user can also opt to calculate the autocorrelation length, which may be taken to be the smallest recommended lag $n$. 
The \texttt{autoCorrelationLength} method  accepts the same inputs as \texttt{autocorrelation} and
returns a single float. Both of these methods make use of the \texttt{emcee} package \cite{emcee}. Here is an example of the usage of both methods.
\begin{lstlisting}[language=Python,numbers=none]
autocorrelations = network.autocorrelation(inputData,
                                           75) #nmax
corrLength = network.autoCorrelationLength(inputData, 
                                           75) #nmax
                                           
\end{lstlisting}


\color{black}

\color{black}
\subsection{Algorithmic Adjustments to Prior Algorithms}
\label{subsec:adjust}
The HMC parameter adaptation scheme briefly mentioned in Section 3.2 is a modified version of the method described in Ref.\,\cite{Wang:2013:AHR:3042817.3043100}. Our method adapts the number of steps and stepsize in a leapfrog trajectory in attempt to maximize the length of the trajectories in the network parameter space in each iteration of the HMC sampler.

In \texttt{TensorBNN}, two variations are introduced. First, in the case that the HMC sampler iterates through  $50$ leapfrog trajectories without having accepted a point during burn-in, \color{black} the algorithm resets and begins with the stepsize maximum and minimum values decreased by a factor of two. This was empirically observed to prevent the HMC sampler from remaining stagnant with an extremely low acceptance rate.
Additionally, after each reset of the parameter adapter, the value of the step size and number of leapfrog steps were randomized for the first 20 iterations in order to prevent the algorithm from converging to a boundary point of the interval that had a high acceptance, but was not optimal, as it was observed to do without this randomization. This algorithm was implemented using a combination of \texttt{TensorFlow} and \texttt{NumPy} \cite{scipy}.

In order to begin the Markov chain sampling of network parameters at a position that reduces the number of needed burn-in steps, we trained a fully connected neural network using \texttt{AMSGrad}\,\cite{j.2018on} with the version of \texttt{Keras} in \texttt{TensorFlow}.  This can be done using \texttt{TensorBNN} or independently by the user should additional control be desired. \color{black}
We observed empirically that choosing a starting point for the Markov chain from a pre-training with gradient descent led to a faster burn-in time. The pre-training is done through three training cycles, each containing a fixed  maximum number of epochs. Each cycle runs until the specified maximum number of epochs or until the validation loss has not decreased for a chosen number of epochs, which we call the patience parameter. \color{black}  After each cycle the learning rate was decreased by a factor of ten, and the best network, judged by the loss computed using the validation data, is selected as the starting point for the next cycle.
\color{black}

\section{Use Cases}
\label{sec:usecases}

\begin{figure}
    
    \centering
    \includegraphics[keepaspectratio, width = 6.5cm]{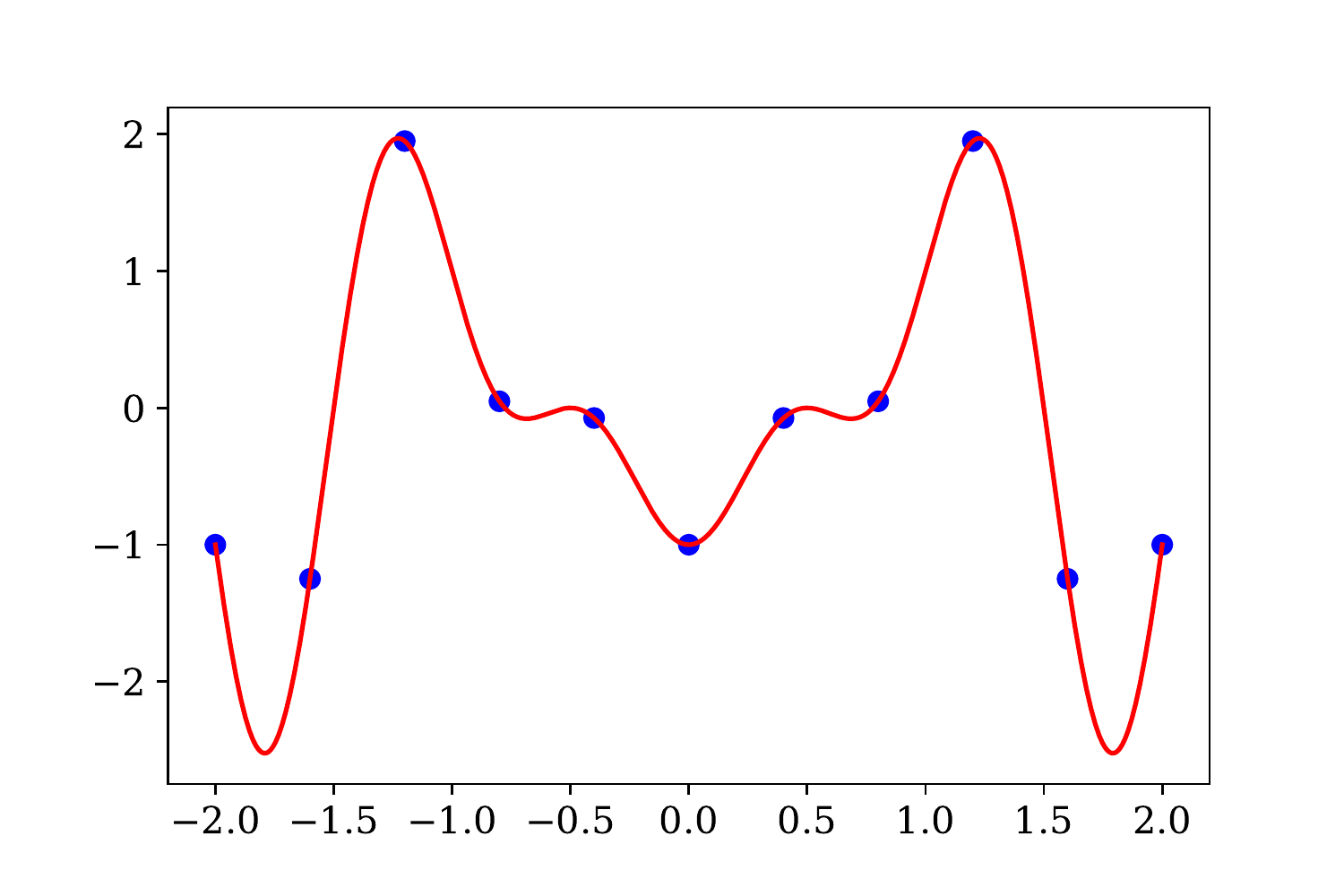}
    \includegraphics[keepaspectratio, width = 6.5cm]{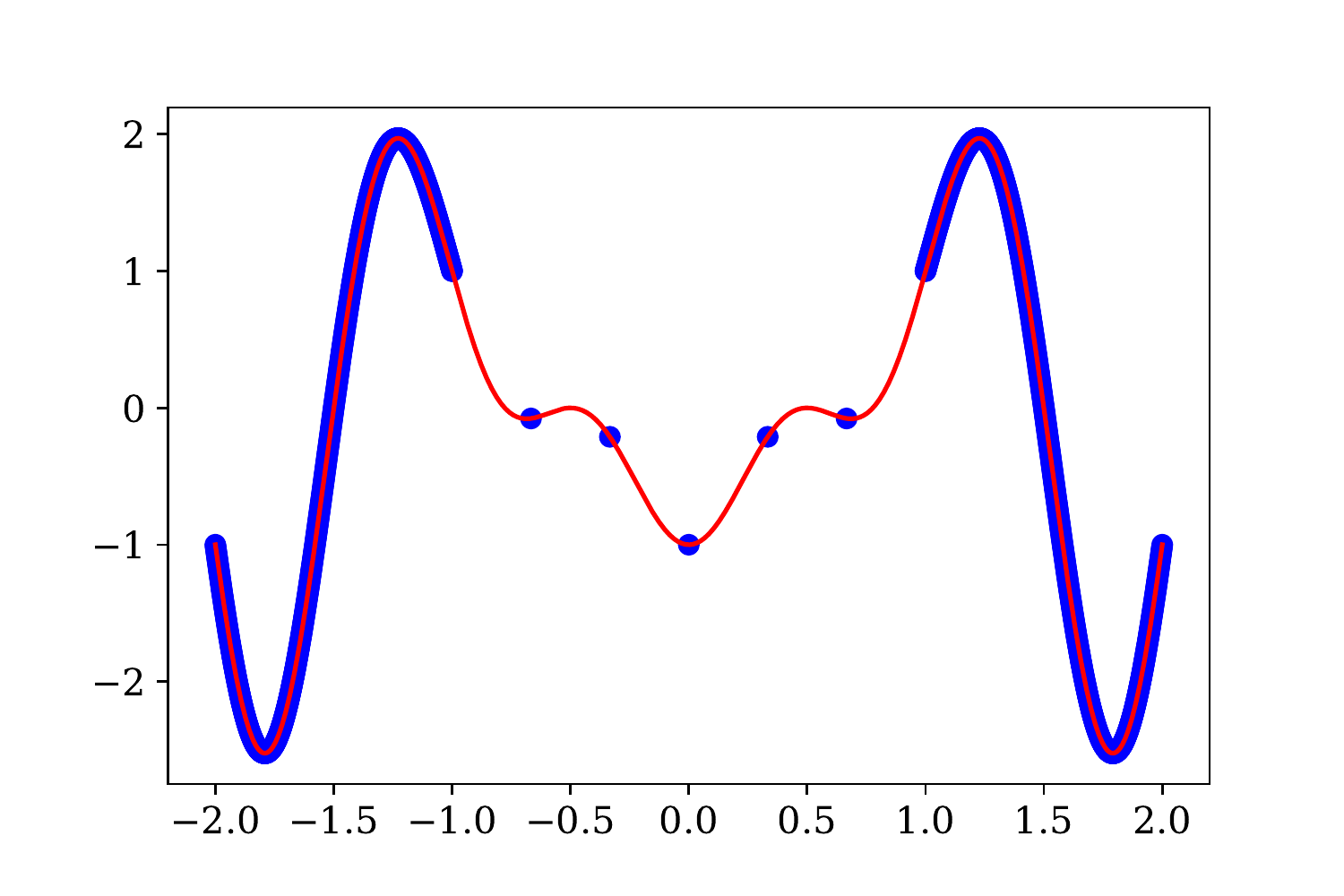}
    \caption{The training data (blue dots) and actual curve (red line) for
    the training examples. The \textbf{sparse dataset} is on the left and \textbf{variable dataset} is on the right}
    \label{fig:input_data}
\end{figure}

Here we present a simple application of \texttt{TensorBNN} in which we train a BNN to learn the function 
\begin{equation}
    y = x\sin(2\pi x)-\cos(\pi x).
\end{equation}
To accomplish this  task  we construct two datasests. One has 11 evenly spaced points $(x_i, y_i)$ for $x_i\in[-2,2],$ called the \textbf{sparse dataset}, and the second, called the {\bf variable dataset}, is the union of the following sets: 1024 points with $x_i$ evenly spaced in $[-2,-1]$, 1024 points with $x_i$ evenly spaced in $[1,2],$ and 7 points with $x_i$ evenly spaced in $[-1,1].$  The training data and true curve are shown in Figure\,\ref{fig:input_data}.  
In addition, we present results demonstrating the performance of a model, trained using gradient descent, for approximating BNNs using current \texttt{TensorFlow} and \texttt{TensorFlow}-\texttt{Probability} functionality. The model uses \texttt{DenseFlipout} layers to approximates the posterior density of network parameters using Gaussian densities and using Gaussian priors. Both models are tested on data with $x_i\in[-3,3].$ The goal is assess the performance of the models on sparse data, dense data, variably distributed data, and extrapolation.

The network architecture for these tests consists of three fully connected hidden layers with 10 nodes each using the hyperbolic tangent activation function. The starting point for sampling was obtained by pre-training the model with the AMSGrad optimizer for 100 epochs with a patience of 10 at learning rates of 0.01, 0.001, and 0.0001. The values for each of the remaining training parameters can be found in Table\,\ref{tab:bounds}. As the dataset used for this task is small and the network is not especially large training can be completed using either a GPU or a CPU in a matter of minutes. The \texttt{TensorFlow}-\texttt{Probability} models are trained with the same network architecture with the AMSGrad optimizer for 1000 epochs with a patience of 100 at learning rates of 0.01, 0.001, and 0.0001.
\begin{table}[ht]

\begin{center}
\begin{tabular}{ | p{3.2cm} | p{1.5cm} | p{2.0cm} |} 
\hline
Parameter & Value (\textbf{sparse}) & Value (\textbf{variable}) \\ 
\hline
leapfrog start & 1000 & 1000 \\ 
\hline
leapfrog min & 100 & 100 \\ 
\hline
leapfrog max & 10000 & 10000 \\ 
\hline
step size start & 1e-3 & 1e-3 \\ 
\hline
step size min & 1e-4 & 1e-4 \\ 
\hline
step size max & 1e-2 & 1e-2 \\ 
\hline
leapfrog grid step & 10 & 10 \\ 
\hline
hyper step size & 1e-3 & 1e-3 \\ 
\hline
hyper leapfrog & 10 & 10 \\ 
\hline
burn-in & 1000 & 1000 \\
\hline
epochs & 6001 & 6001 \\ 
\hline
trainable hyper-priors & True & True \\ 
\hline
trainable likelihood sd & False & False \\ 
\hline
\end{tabular}
\end{center}
\caption{network parameters}
\label{tab:bounds}
\color{black}
\end{table}

\begin{figure}

\begin{tabular}{p{6.0cm}p{6.0cm}}
  \includegraphics[keepaspectratio, width = 6.0cm]{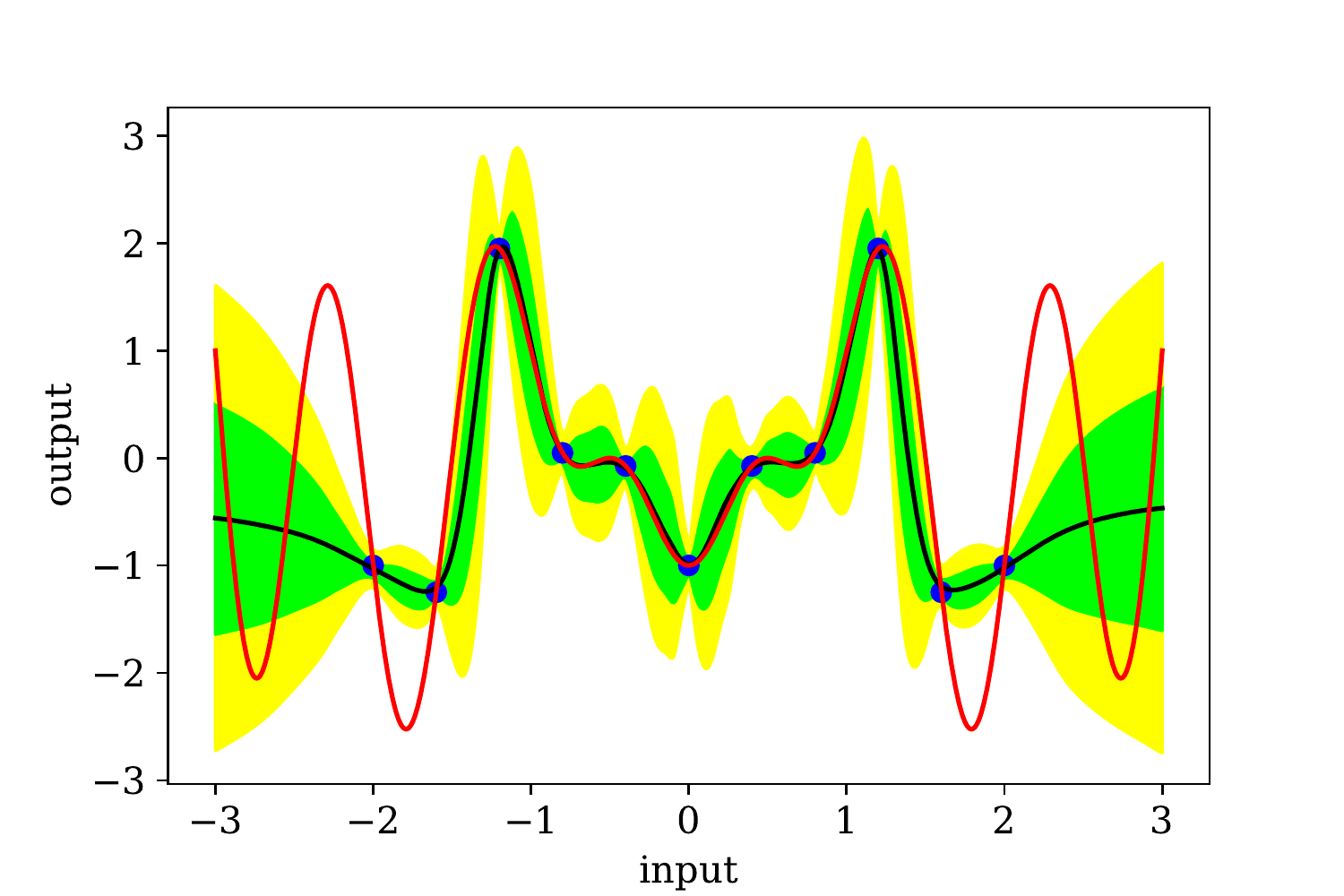} &   \includegraphics[keepaspectratio, width = 6.0cm]{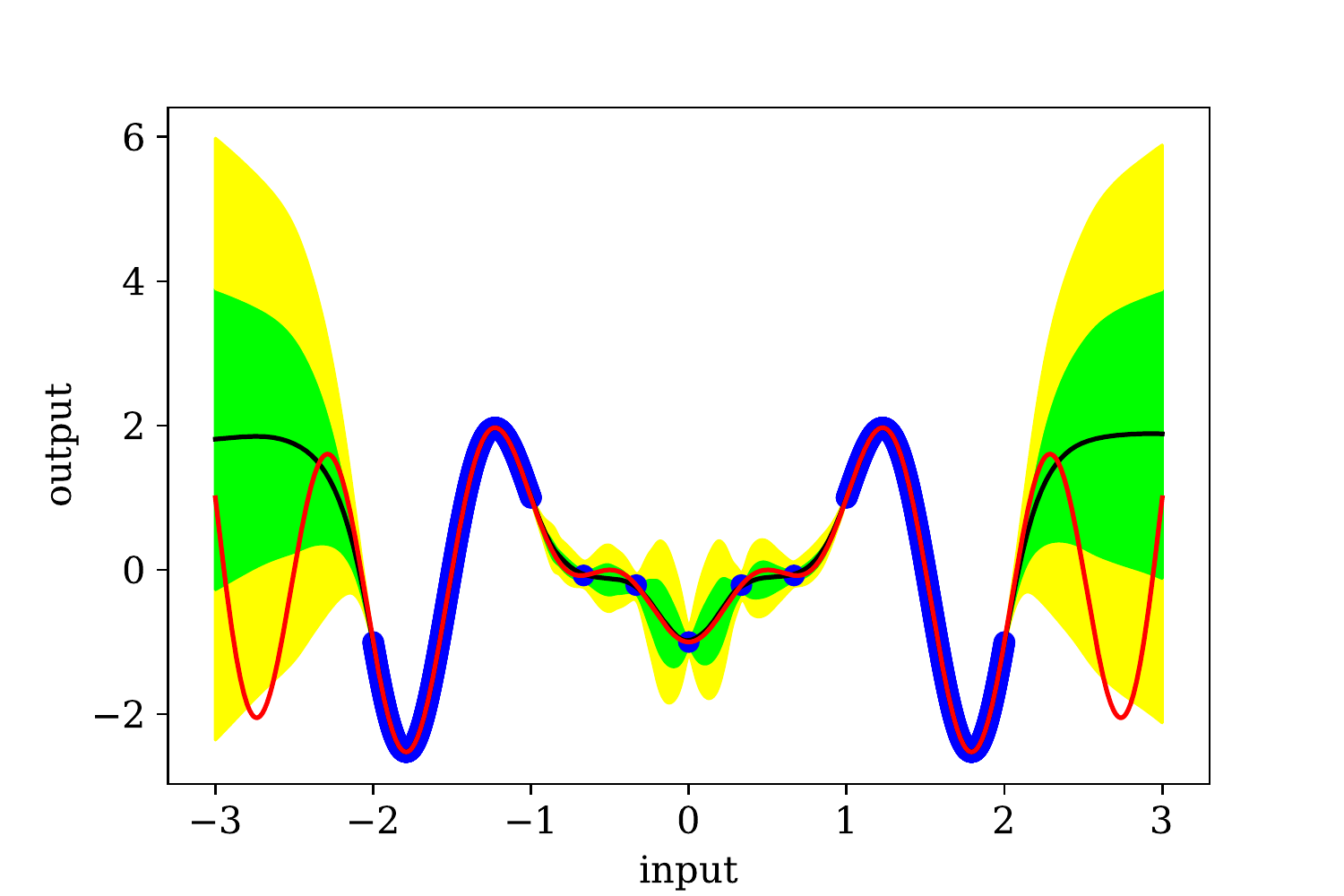}

\end{tabular}
\caption{The training data (blue dots) and true curve (red line) for the \textbf{sparse dataset} (left) and \textbf{variable dataset} (right), along with the mean of the predictions from \texttt{TensorBNN} (black line), and the mean plus and minus 1 and 2 standard deviations (green and yellow shading).
}
    \label{fig:BrazilTBNN}
\end{figure}

A graphical representation of the training results for using \texttt{TensorBNN} is shown in Figure\,\ref{fig:BrazilTBNN}. These plots demonstrate several important properties of BNN. Firstly, we see generally that as the test points get further from the training points the variability of the output of the BNN increases\color{black}. 
\color{black} This is consistent with the fact that there are more predictions that are consistent with the training data, while the variability is small near the training points as expected and desired.

\begin{figure}

\begin{tabular}{p{6.0cm}p{6.0cm}}

  \includegraphics[keepaspectratio, width = 6.0cm]{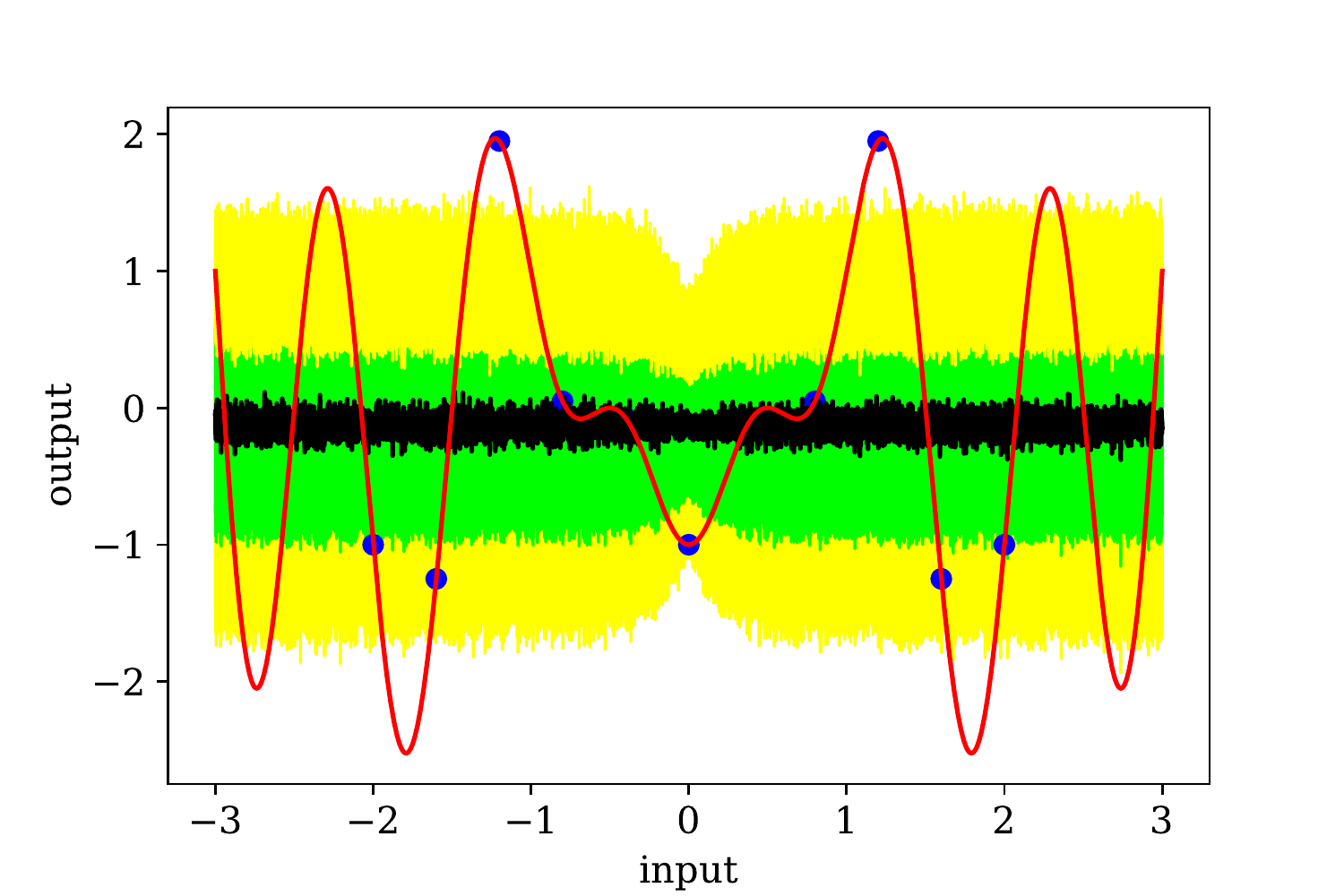} &   \includegraphics[keepaspectratio, width = 6.0cm]{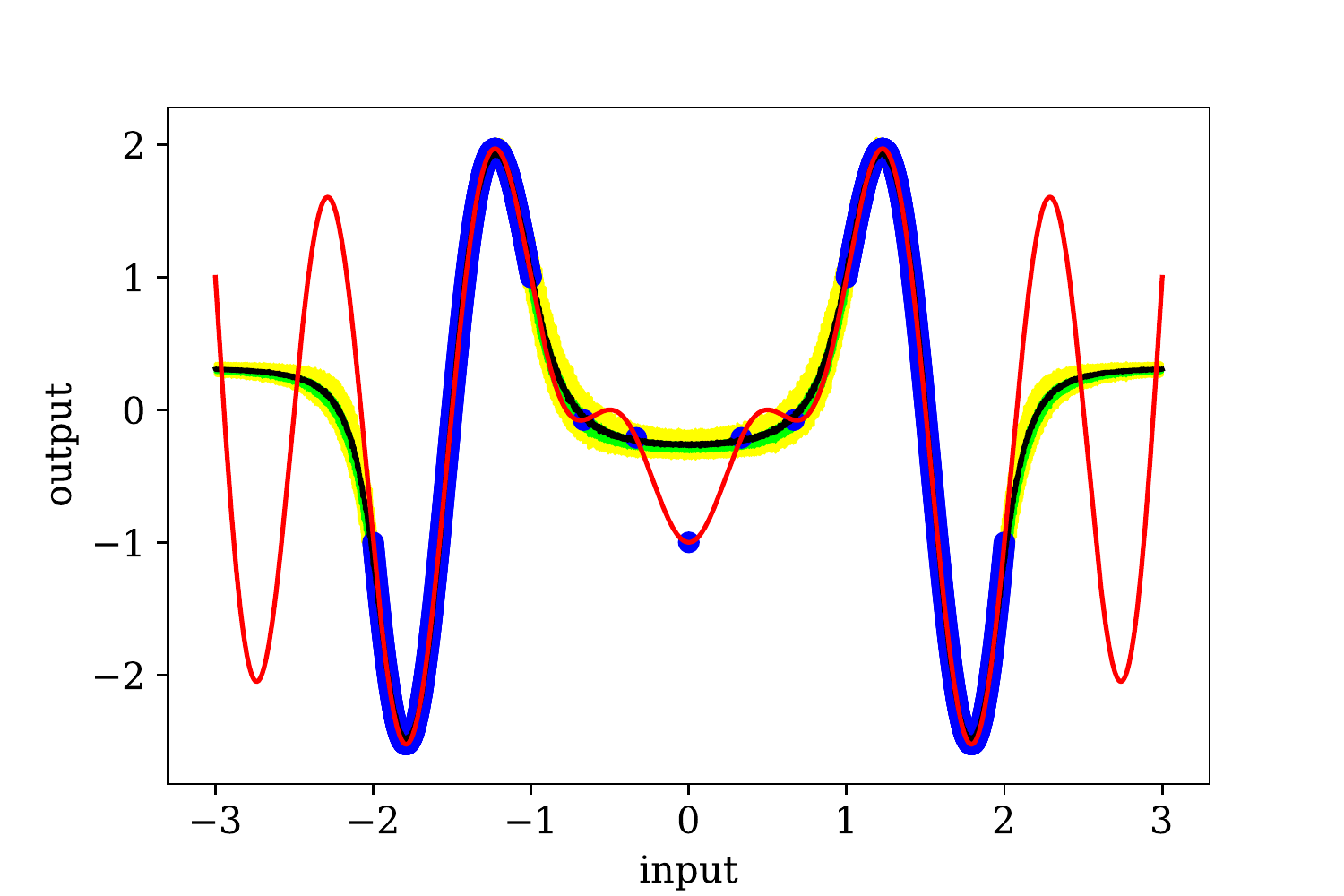} \\

\end{tabular}
\caption{The training data (blue dots) and true curve (red line) for the \textbf{sparse dataset} (left) and \textbf{variable dataset} (right), along with the mean of the predictions (black line) from \texttt{TensorFlow}-\texttt{Probability}, and the mean plus and minus 1 and 2 standard deviations (green and yellow shading). 
}
    \label{fig:BrazilTFP}
\end{figure}

Secondly, the credible intervals of the BNN do not always contain the true value. In the left graph in Figure\,\ref{fig:BrazilTBNN}, it is clear that the BNN is quite wrong about the behavior of the curve between the first two and last two training points. This, however, is not unreasonable, as the value of the curve decreases dramatically between those values, without any training data to indicate this behavior. 
 The BNN predicts that the curve should follow a roughly straight path between the points, 
but correctly suggests that the prediction in that region is very uncertain.
But while the BNN will predict greater uncertainty in regions without data, as desired, it remains true, nevertheless, that one should be cautious about any extrapolation into a region with no data because the predicted uncertainty will depend on details of the model, in particular, the prior. 
It is therefore good practice to study the sensitivity of conclusions to modifications of the prior in any real-world analysis. {\tt TensorBNN} provides tools for such sensitivity studies.

An interesting approximation to the posterior density of neural network parameters, and therefore to Bayesian neural networks,  is available through the \texttt{DenseFlipout} and \texttt{DenseVariational} layers of  \texttt{TFP} \cite{tfp}. This method approximates a Bayesian neural network using a combination of variational inference and sampling. A forward pass through the network samples the network parameters from a variational density, a diagonal multivariate Gaussian whose  means and standard deviations are optimized during training, a method originally proposed in \cite{NIPS2011_7eb3c8be}. 

\texttt{TensorBNN} on the other hand makes no assumption about the form of the posterior density and, therefore, in principle can account for posterior densities containing non-trivial dependencies between the network parameters.
Given the fundamentally different approximations used in \texttt{TensorBNN} and \texttt{TensorFlow-Probability}, {\it a priori}, it is not be surprising if these approximations sometimes yielded different predictive distributions. 
A systematic comparison of the two approaches is beyond the scope of this paper, but in Figure\,\ref{fig:BrazilTFP} we see an example of how the results can differ. Figure\,\ref{fig:BrazilTFP} shows the results of our toy problems using \texttt{DenseFlipout} layers, which may be compared with the results obtained using \texttt{TensorBNN} in Figure\,\ref{fig:BrazilTBNN}. The results using the different approaches are reasonable under the respective assumptions including the choice of priors. It would be interesting to investigate whether the results of the two approaches would converge as the priors permitted a greater and greater dynamic range in the values of the network parameters. Additionally, BNNs in the style of those in \texttt{TensorFlow-Probability} have been used successfully in HEP, such as in \cite{plehn1, plehn2}.

\color{black}

Beyond these simple examples, a more complex application of \texttt{TensorBNN} can be found in \cite{kronheimSUSY}. In that paper, the package is used to predict the cross sections for supersymmetric particle creation at the Large Hadron Collider, predict which supersymmetric model points are theoretically viable, and predict the mass of the lightest neural Higgs boson, which is generally identified with the Higgs boson discovered at CERN\,\cite{Aad:2012tfa,Chatrchyan:2012ufa}.

\section{Summary}
\label{sec:conclusion}
 \texttt{TensorBNN} is a framework that allows for the full Bayesian treatment of neural networks while leveraging GPU computing through the \texttt{TensorFlow} platform. Through an automation of the search for the parameters needed for an effective Hamiltonian Monte Carlo sampler and the ease of using a pre-trained network, \texttt{TensorBNN} is able to decrease the computation needed to converge to a sample from the posterior density of the neural network parameters. The algorithm automatically adapts hyper-parameters to reduce the amount of fine-tuning required by the user. As shown through the simple examples, the distribution of the network predictions behave according to expectations and can be yield good results even on difficult learning problems. The package provides a flexible 
 means to perform regression and binary classification and is designed with the potential for expansion in terms of allowed network architectures and output possibilities.
 
\section{Acknowledgements} 
We would like to thank the \texttt{TensorFlow}-\texttt{Probability} group at Google Research for fruitful discussions surrounding prior choice, hyper-parameter tuning, and re-scaling.
This work was supported in part by the Davidson Research Initiative and the U.S. Department of Energy Award No. DE-SC0010102.






\bibliographystyle{elsarticle-num-names}
\bibliography{main.bib}







\end{document}